\documentclass[aps,prc,twocolumn,showpacs,floatfix,preprintnumbers,superscriptaddress,amsmath,amssymb,longbibliography]{revtex4-1}

\usepackage{graphicx}
\usepackage{dcolumn}
\usepackage{hyperref}
\usepackage{bm}

\newcommand{\rcsection}[1]{{\it #1.~}\nolinebreak}

\newcommand{\co}{(Color online)}

\newcommand{\bra}[1]{\left\langle #1 \right|}
\newcommand{\ket}[1]{\left| #1 \right\rangle}

\newcommand{\NNopt}{NNLO$_{\rm opt}$}

\newcommand{\Leff}{\ensuremath{L_{\rm eff}}}

\newcommand{\Lameff}{\ensuremath{\Lambda_{\rm eff}}}

\newcommand{\nmax}{\ensuremath{N^{\rm tot}_{\rm max}}}

\newcommand{\nmaxtot}{\ensuremath{N^{\rm tot}_{\rm max}}}
\newcommand{\nmaxncsm}{\ensuremath{N^{\rm NCSM}_{\rm max}}}
\newcommand{\hw}{\ensuremath{\hbar\omega}}

\newcommand{\scL}{\ensuremath{\mathcal{L}}}

\newcommand{\ncsm}{NCSM}

\newcommand{\Ei}{\ensuremath{E_{\infty}}}

\newcommand{\be}{\begin{equation}}
\newcommand{\ee}{\end{equation}}
\newcommand{\bea}{\begin{eqnarray}}
\newcommand{\eea}{\end{eqnarray}}

  \def\nuc#1#2{\relax\ifmmode{}^{#1}{\protect\text{#2}}\else${}^{#1}$#2\fi}
  \def\itnuc#1#2{\setbox\@tempboxa=\hbox{\scriptsize\it #1}
    \def\@tempa{{}^{\box\@tempboxa}\!\protect\text{\it #2}}\relax
    \ifmmode \@tempa \else $\@tempa$\fi}
 
\begin{document}

\title{Infrared length scale and  extrapolations for the no-core shell model}

\author{K.~A.~Wendt} 
\affiliation{Department of Physics and Astronomy, University of
   Tennessee, Knoxville, Tennessee 37996, USA}
\affiliation{Physics Division, Oak Ridge National Laboratory, Oak
   Ridge, Tennessee 37831, USA}

\author{C.~Forss\'en} 
\affiliation{Department of Fundamental Physics,
  Chalmers University of Technology, SE-412 96 G\"oteborg, Sweden}
\affiliation{Department of Physics and Astronomy, University of
  Tennessee, Knoxville, Tennessee 37996, USA} 
\affiliation{Physics Division, Oak Ridge National Laboratory, Oak
  Ridge, Tennessee 37831, USA}

\author{T.~Papenbrock} 
\affiliation{Department of Physics and
  Astronomy, University of Tennessee, Knoxville, Tennessee 37996, USA}
\affiliation{Physics Division, Oak Ridge National Laboratory, Oak
  Ridge, Tennessee 37831, USA}

\author{D.~S\"a\"af} 
\affiliation{Department of Fundamental Physics,
  Chalmers University of Technology, SE-412 96 G\"oteborg, Sweden}

\begin{abstract}
  We precisely determine the infrared (IR) length scale of the no-core
  shell model (\ncsm{}). In the NCSM, the $A$-body Hilbert space is
  truncated by the total energy, and the IR length can be determined
  by equating the intrinsic kinetic energy of $A$ nucleons in the
  \ncsm{} space to that of $A$ nucleons in a $3(A-1)$-dimensional
  hyper-radial well with a Dirichlet boundary condition for the
  hyper radius $\rho$. We demonstrate that this procedure indeed
  yields a very precise IR length by performing large-scale \ncsm{}
  calculations for \nuc{6}{Li}. We apply our result and perform
  accurate IR extrapolations for bound states of \nuc{4}{He},
  \nuc{6}{He}, \nuc{6}{Li}, and \nuc{7}{Li}. We also attempt to
  extrapolate \ncsm{} results for \nuc{10}{B} and \nuc{16}{O} with
  bare interactions from chiral effective field theory over tens of
  MeV.
\end{abstract}


\maketitle

\rcsection{Introduction} The spherical harmonic oscillator basis is a
convenient and popular choice in nuclear structure calculations because it
reflects the symmetries and the self-bound character of atomic
nuclei. A finite oscillator space, defined by a maximum of $N$ excited
oscillator quanta and frequency $\omega$, exhibits infrared (IR) and
ultraviolet (UV) cutoffs $\pi/ L$ and $\Lambda$,
respectively~\cite{stetcu2007}.  Here, $L\approx \sqrt{2N} b$ and
$\Lambda\approx\sqrt{2N}/b$ (in units where $\hbar=1=c$) are leading-order (LO) approximations in
$N$, valid for $N\gg 1$~\cite{hagen2010b,jurgenson2011}, and $b\equiv
\sqrt{\hbar/(M\omega)}$ denotes the oscillator length for a particle
of mass $M$.  This makes it necessary to understand the convergence of
energies and other observables as $L$ and $\Lambda$ are increased. The
UV convergence depends on the momentum regulators employed in the
nuclear interaction~\cite{koenig2014}, while the IR convergence
depends on the structure of the nucleus under
consideration. \textcite{coon2012} found that the IR convergence of
ground-state energies is exponential in $L$ (in model spaces where
corrections due to a finite UV cutoff $\Lambda$ can be
neglected). This exponential convergence can be understood as follows
\cite{furnstahl2012}: For long wavelengths, the finite oscillator
basis is indistinguishable from a spherical well with a hard wall at
a radius $L$, and the resulting Dirichlet boundary condition induces
corresponding corrections to the exponential fall-off of bound-state
wave functions. This insight allows one to derive IR extrapolation
formulas for bound-state energies and
radii~\cite{furnstahl2012,furnstahl2014}.

For IR extrapolations to work in practice, one needs a value for the IR length
$L$ that is more precise than the LO result given in the previous paragraph.
As it turns out, the next-to-leading order (NLO) value of $L$ depends on the
model space employed in the calculation, but the method to 
compute the IR length is system independent. For a single particle in
$d=3$ dimensions (or the deuteron in the center-of-mass system),
\textcite{more2013} derived a very precise value of $L$ by equating the lowest
eigenvalue of the squared momentum operator in the finite oscillator basis with
$(\pi/L)^2$, i.e. the lowest eigenvalue of the squared momentum operator in the
infinite spherical well of radius $L$. The result (for a single particle in $d$
dimensions) is
\begin{equation}
L=L_2(d)\equiv\sqrt{2(N+d/2+2)}b.
\label{eq:L2}
\end{equation} 
This result is NLO in $N$. While derived for a single-particle system, it has also been
applied in extrapolations of nuclei with mass numbers $A>2$; see, e.g.,
Refs.~\cite{soma2013,jurgenson2013,saaf2013,roth2013}.

Very recently, \textcite{furnstahl2014b} derived a precise value of the IR length scale 
for $A$-fermion systems whose Hilbert space is a Cartesian product of
single-particle oscillator spaces truncated at $N$. Such a Hilbert
space is employed by several quantum many-body
methods~\cite{dickhoff2004,dean2004,hagen2008,tsukiyama2011,hagen2013c,hergert2013b,binder2013b,soma2013}. The
key was again to equate the lowest eigenvalue of the total squared
momentum operator in the finite oscillator basis to the lowest
eigenvalue of the $A$-body kinetic energy in an infinite spherical well of
radius $L$, keeping the exact dependence on $N$.  The resulting IR
length $L$ differs in NLO from $L_2$, and numerical values are tabulated in
Ref.~\cite{furnstahl2014b}.

We are still lacking a precise value of the IR length scale $L$ for the
many-body model space truncation employed in the no-core shell model
(\ncsm{})~\cite{navratil2009,barrett2013}. This widely used
method~\cite{navratil2000,forssen2005,quaglioni2008,
roth2008a,dytrych2008,maris2009,johnson2013} employs a total energy truncation,
i.e. a Hilbert space of all $A$-body product states with an energy not exceeding
$\nmaxtot\hbar\omega$. In the \ncsm{} literature, the model space is usually
specified by the number of excitations above the lowest configuration for the
symmetry (parity, numbers of protons and neutrons) of interest. We will denote
this truncation by \nmaxncsm{} in order to distinguish it from the total number
of \hw{} quanta, \nmaxtot{}.  The many-body character of this truncation implies
that the total squared momentum operator is not a single-particle operator in
this model space (and its eigenstates are not product states). Thus, the IR
scale derived by \textcite{furnstahl2014b} is only a leading order approximation
of the many-body IR length scale.  It is the purpose of this Rapid Communication
to precisely determine the IR length scale of the NCSM.

We finally note that the convergence and corrections due to finite model spaces
are also studied for interacting particles on lattices. Here, too, the effects
of hard walls or periodic boundary conditions onto many-body bound states are of
particular interest~\cite{luscher1985,koenig2011,koenig2012,pine2013,briceno2013,briceno2013b}. In contrast to the harmonic oscillator, the precise IR and UV
cutoffs are easily identified on the lattice, and the effort goes into
extrapolation formulas for relevant observables. 

\rcsection{Infrared length scale of the \ncsm{}} 
Let us consider $A=3$ spinless fermions in $d=1$ dimensions as an
illustrative example. Following
Refs.~\cite{more2013,furnstahl2014,furnstahl2014b}, we seek to equate
the kinetic energy of this system in the NCSM space to the
kinetic energy of a corresponding system in an infinite well
of radius $L$. Our task consists of determining what the corresponding
system really is. The Hilbert space is spanned by Slater determinants
\be
 \phi_{n_1 n_2 n_3}(x_1,x_2,x_3) = {\rm det}\left[\psi_{n_i}(x_j)\right]_{i,j=1,2,3}
\ee
of harmonic oscillator (HO) wave functions $\psi_n(x)$, and we only include three-body states
that fulfill the total energy truncation
\begin{align}
\label{eqn:Ntot}
  \sum_{i=1}^A n_i \le \nmaxtot\,.
\end{align}
The key insight is that this Hilbert space of $A=3$ particles in $d=1$
dimensions is equivalent to that of a single particle in $Ad=3$ dimensions,
spanned by three-dimensional spherical harmonic oscillator wave functions
\be
  \phi_{n l m}(\mathbf{r})=\phi_{n l m}(r, \theta, \varphi) = 
      R_{n l}(r) Y_{lm}(\theta, \varphi)\,.
\ee
Here, $R_{nl}$ is a radial wave function and $Y_{lm}$ are spherical harmonics.
The NCSM truncation of Eq.~\eqref{eqn:Ntot} is equivalent to allowing only those
single-particle basis states $\phi_{n l m}$ with $2n+l\le\nmax$. However, we
also need to consider the antisymmetry of the $A=3$ wave function. If we align
the projection axis of the spherical basis along the line $x_1=x_2=x_3$,
antisymmetry can be obtained with wavefunctions proportional to $\sin{3\varphi}$
, i.e., $m$ needs to be a multiple of three, which implies $l\ge3$. Thus, with
this additional symmetry constraint, the NCSM truncation for $A=3$ particles in
$d=1$ dimensions naturally corresponds to a single particle in $Ad=3$
dimensions, with single-particle energies limited to $\nmaxtot{}$. The IR
properties of the single particle in a three-dimensional oscillator space are
well known~\cite{more2013}, and the harmonic oscillator truncation imposes a
Dirichlet-like boundary condition on the radial coordinate.

\begin{figure}[htbp]
  \centering
  \includegraphics[width=3.4in]{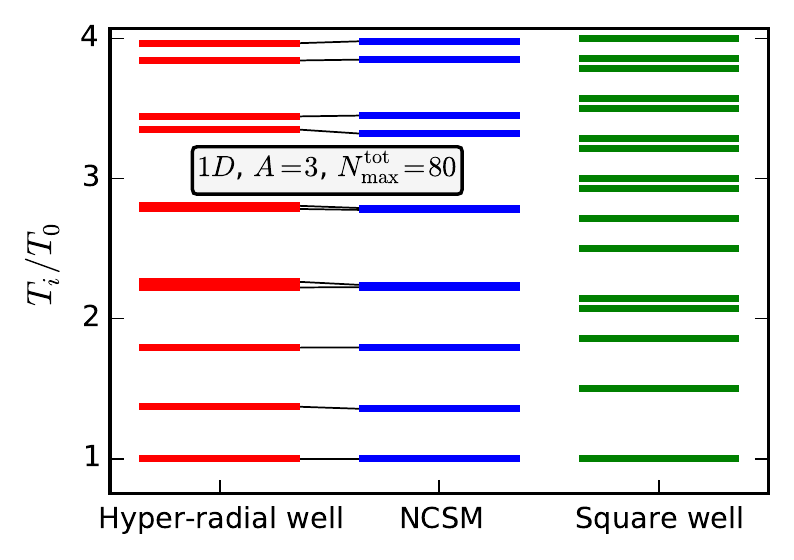}
  \caption{\co{} Comparison of kinetic energy spectra for three fermions in a hyper-radial 
  well (left), three fermions in a one-dimensional NCSM basis (middle), and three 
  fermions in a one-dimensional infinite square well (right). }
  \label{fig:OneDKE}
\end{figure}

As a check, we compute the eigenvalues of the kinetic energy for $A=3$
fermions in $d=1$ dimension (in a NCSM model space with $\nmaxtot=80$)
and compare them to the kinetic-energy spectrum of a three-dimensional
hyper-radial well.  The antisymmetry of the former system manifests
itself as a discrete symmetry of the latter.  The results are shown in
Fig.~\ref{fig:OneDKE} (middle and left spectrum, respectively) and
also compared to the kinetic energy spectrum for three fermions in a
one-dimensional infinite well (right spectrum).  In each case, the
entire spectrum is proportional to the inverse square of an underlying
length scale, so we plot the eigenvalues $T_i$ in units of the lowest
kinetic energy eigenvalue $T_0$ to remove this dependence.  Clearly
the \ncsm{} spectrum closely matches that of the hyper-radial well but not
that of three particles in an infinite square well.

We can generalize these results as follows. The Hilbert space of
$A$ nucleons in $d=3$ dimensions subject to the NCSM truncation
identified by $\nmaxtot$ is equivalent to that of a single particle,
with certain discrete symmetry constraints, in an $Ad$-dimensional HO space
with single-particle energies up to $\nmaxtot$ excited quanta. At low
momenta, the latter is equivalent to a hyper-radial
well. Equating the kinetic energies yields the size of this well and
consequently the IR length $L$ of the corresponding \ncsm{} basis. Alternatively, 
the NCSM truncation can also be viewed as a system of $A$ fermions confined to an 
$Ad$ dimensional hyper-radial well. 

Let us therefore compute the eigenvalues of the kinetic energy for a
$D$-dimensional hyper-radial well with an infinite wall at hyper radius
$L$. The hyperspherical basis states can be labeled as
$\ket{\rho\,G\,\bm{\alpha}}$, where $\rho$ is the hyper radius, $G$ is
the grand angular momentum, and $\bm{\alpha}$ is the collection
of all other partial-wave quantum numbers.  The kinetic energy
operator is block diagonal in both $G$ and $\bm{\alpha}$, so we
focus on a single arbitrary hyperspherical partial wave.
The hyper-radial part of the noninteracting Hamiltonian is
\bea
  -\left(
    \frac{\partial^2}{\partial\rho^2} 
    -
    \frac{\scL(\scL+1)}{\rho^2}
  \right)\psi_{G}(\rho)
  =
  Q^2 \psi_{G}(\rho)\,,
\eea
where $\scL = G + (D-3)/{2}$ and $Q^2$ is the total squared momentum. The
hyper-radial eigensolutions of this Hamiltonian are
\be
  \psi_{G}(\rho)  = \sqrt{Q \rho} J_{\scL+\frac{1}{2}}(Q\rho)\,,
\ee
where $J_{\nu}(X)$ is a Bessel function of the first kind.  Imposing a Dirichlet
boundary condition at $\rho=L$ implies that $QL$ is a zero of
$J_{\scL+\frac{1}{2}}$. We denote the $i$th zero as $X_{i,\scL}$.  The
selection criteria for $\scL$ (yielding an antisymmetric wave function)
is discussed below, but the entire spectrum of our well is now completely
determined by a minimum value of $\scL$ and the hyper radius $L$ of the well
\be
  \{Q^2_{i,n}\} = \{L^{-2}X^2_{i,\scL_{\rm min}+2n} \;\forall i,n \in
  \mathbb{Z}\,\}.
  \label{eq:Thyperradial}
\ee
Here $\scL=\scL_{{\rm min} +2n}$ labels states of the same parity.

The next critical ingredient is the lowest eigenvalue of the kinetic energy
operator in the \ncsm{} basis.  Recall that even though $\hat{T}$ is a one-body
operator, the NCSM  truncation effectively promotes it to an $A$-body operator
\begin{multline}
  \hat{T}_\textrm{\ncsm{}} =
      \left(\sum^{A}_{i=1}\hat{p}^2_{i}\right)\\
      \times \Theta\left(\nmaxtot \hw -
        \sum^{A}_{i=1}\left({\hat{p}^2_i\over 2M} +{M\omega^2\over
            2}\hat{x}^2_i-{3\over 2}\hw\right)\right)\,. 
\end{multline}
Here $\Theta$ denotes the unit step function that enforces the \ncsm{}
truncation. Even though  $\hat{T}_\textrm{\ncsm{}}$ is an $A$-body operator, the
hyperspherical basis can be used to ease the computational requirement for
finding its eigenvalues.  Similar to the example discussed above, we can expand
any product of three-dimensional HO states into hyper-radial harmonic oscillator
states. 

Likewise the transformation is block diagonal in the total oscillator
quanta $\sum^{A}_{i=1} (2n_i+l_i) = 2N + G$, where $N$ is the nodal quantum
number for the hyper-radial coordinate. Exploiting this block diagonal
structure, we need to only diagonalize small matrices, with dimension $5$ --
$20$, instead of the full dimension of the \ncsm{} basis.  The kinetic energy
matrix elements are
\bea
  \lefteqn{\bra{NG\bm{\alpha}} \hat{T}_{\rm\ncsm} \ket{N'G'\bm{\alpha}'}
    =}\nonumber\\
    &&\delta_{G}^{G'}\delta_{\bm{\alpha}}^{\bm{\alpha}'}\frac{\hw}{2}
    \Bigg[
      \delta_{N}^{N'} \left(2N+\scL+\frac{3}{2}\right)\nonumber\\
      &&+ \delta_{N_>}^{N_<+1} \sqrt{
    \left(N_<+1\right)
    \left(N_<+\scL+\frac{3}{2}\right)
    }
    \Bigg]\, ,
\eea
and it will be sufficient to consider a single hyperspherical channel
with grand angular momentum $G$. 
Here $N_{<} \equiv \min{(N,N')}$, $N_> \equiv\max{(N,N')}$, and $N,N'$ run from
0 to $\left\lfloor\frac{\nmax-G}{2}\right\rfloor$, with the brackets $\lfloor
.\rfloor$ denoting the integer part of their argument.  We denote the needed
dimensionless eigenvalues as $T_{i,\scL}(\nmaxtot)$ such that
\begin{equation}
  \hat{T}_{\rm \ncsm} \ket{i} = \frac{\hw}{2}T_{i,\scL}(\nmaxtot) \ket{i}
  \label{eq:Tncsm}
\end{equation}

The smallest permitted eigenvalue is driven by the smallest symmetry-allowed
value of $\scL=G+(D-3)/{2}$. For a single product state, $D=3A$, and $G$ can be
decomposed as
\begin{align}
  G &= \sum_{i=1}^{A} l_i + \sum_{i=1}^{A-1} n_{i,i+i}\,.
\end{align}
Here $l_i$ is the orbital angular momentum and $n_{i,i+1}$ is the nodal quantum
number for the hyper-angle between the radial coordinates $r_i$ and $r_{i+1}$,
and $A$ is the number of single-particle coordinates.  In
a single-particle basis, this means that the lowest possible value for
$G$ is $G_{\rm min,sp} = \sum_i l_{i,0}$, where
$l_{i,0}$ are the orbital quantum numbers from the lowest
(symmetry-allowed) energy configuration in the basis. 

\ncsm{} calculations for $A>6$ usually employ single-particle
coordinates (instead of relative coordinates~\cite{barnea1997,bacca2002}).
However, the \ncsm{} eigenstates are products of a center-of-mass state and an
intrinsic state.  Thus, the relevant IR length is an intrinsic scale. The
dimension of the intrinsic basis is $D=3(A-1)$, and $G_{\min}$ (the
lowest value of the grand angular momentum in the relative coordinate
system) is determined by
the sum of intrinsic orbital angular momenta that can couple with spins to give
the ground-state angular momentum $J$ and parity $\Pi$.  This means
that $G_{\min} = G_{\rm min,sp}$ because the center-of-mass state
carries no angular momentum.

\begin{figure}[htbp]
  \centering
  \includegraphics[width=3.4in]{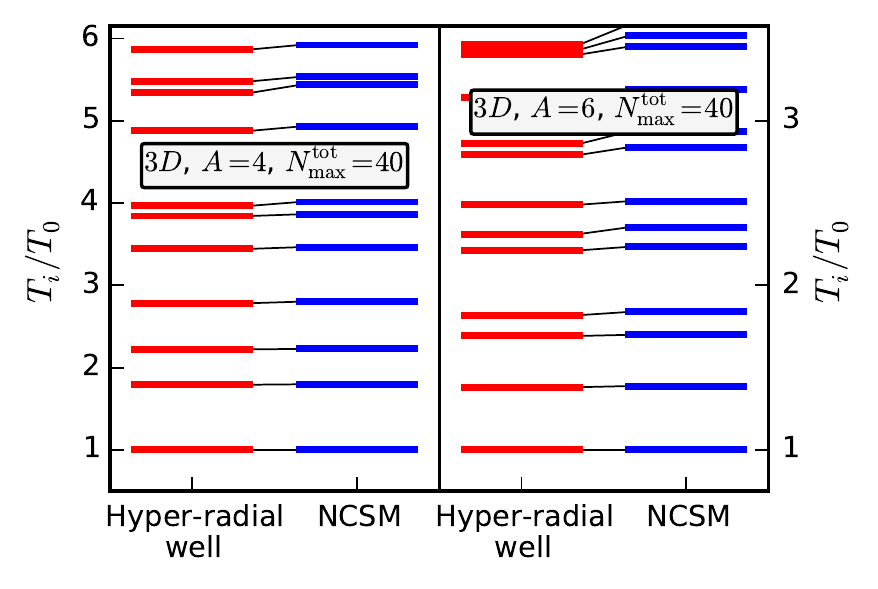}
  \caption{\co{} Discrete intrinsic kinetic energy spectra for $A=4,6$ particles in the \ncsm{}
  (right pane) and for the corresponding $D=3(A-1)$ dimensional hyper-radial
  infinite well.  In each case, we plot in units of the smallest eigenvalue.}
  \label{fig:ThreeDKE}
\end{figure}

The close similarity of the kinetic energy spectra of $3(A-1)$-dimensional
hyper-radial wells with Dirichlet boundary condition and the corresponding
intrinsic kinetic energies in a \ncsm{} basis for $A=4,6$ particles in three
dimensions is shown in Fig.~\ref{fig:ThreeDKE}. For the large values of
\nmaxtot{} employed in this numerical comparison, the agreement between the
spectra persists up to highly excited states.

The intrinsic IR length is now obtained by equating the lowest kinetic energy
eigenstate in the hyper-radial well, from Eq.~\eqref{eq:Thyperradial},
and the first eigenstate in the NCSM basis, from Eq.~\eqref{eq:Tncsm}. 
This yields
\begin{equation}
  L_{\rm eff} = 
      b \frac{X_{1,\scL}}{\sqrt{T_{1,\scL}(\nmax)}}\,
      \label{eq:Leff}
\end{equation}
with
\begin{equation}
  \scL = 
  G_{\min} + 
        \frac{3(A-2)}{2}.
      \label{eq:mathcalL}
\end{equation}
Numerical values for $L_{\rm eff}$ are tabulated in the Supplemental
Material~\footnote{See Supplemental Material at \url{URL will be
    inserted by publisher} for a tabulation of numerical values for
  \Leff{}.}. 

Following~\textcite{koenig2014} we exploit the duality of the HO Hamiltonian
under the exchange of position and momentum operators and identify the UV scale
of the
\ncsm{} as
\begin{align}
  \Lambda_{\rm eff} 
    &= 
      \frac{X_{1,\scL}}{b\sqrt{T_{1,\scL}(\nmax)}}\,
      = L_{\rm eff}/b^2
      \label{eq:Lameff}
\end{align}

To illustrate that \Leff{} is indeed the correct IR scale of
the \ncsm{} basis we perform large-scale calculations
of \nuc{6}{Li} for a wide range of HO frequencies
\hw{}. We used the nucleon-nucleon interaction
\NNopt{}~\cite{ekstrom2013} in model spaces up to $\nmaxncsm=18$
($\nmaxtot=20$). The UV-regulator cutoff of this interaction is
$500$~MeV. The model-space parameters $(\hw,\nmaxncsm)$ were converted
to ($\Leff,\Lambda_\mathrm{eff}$), using Eqs.~\eqref{eq:Leff} and~\eqref{eq:Lameff}.

\begin{figure*}[htbp]
  \centering
  \includegraphics[width=.8\textwidth]{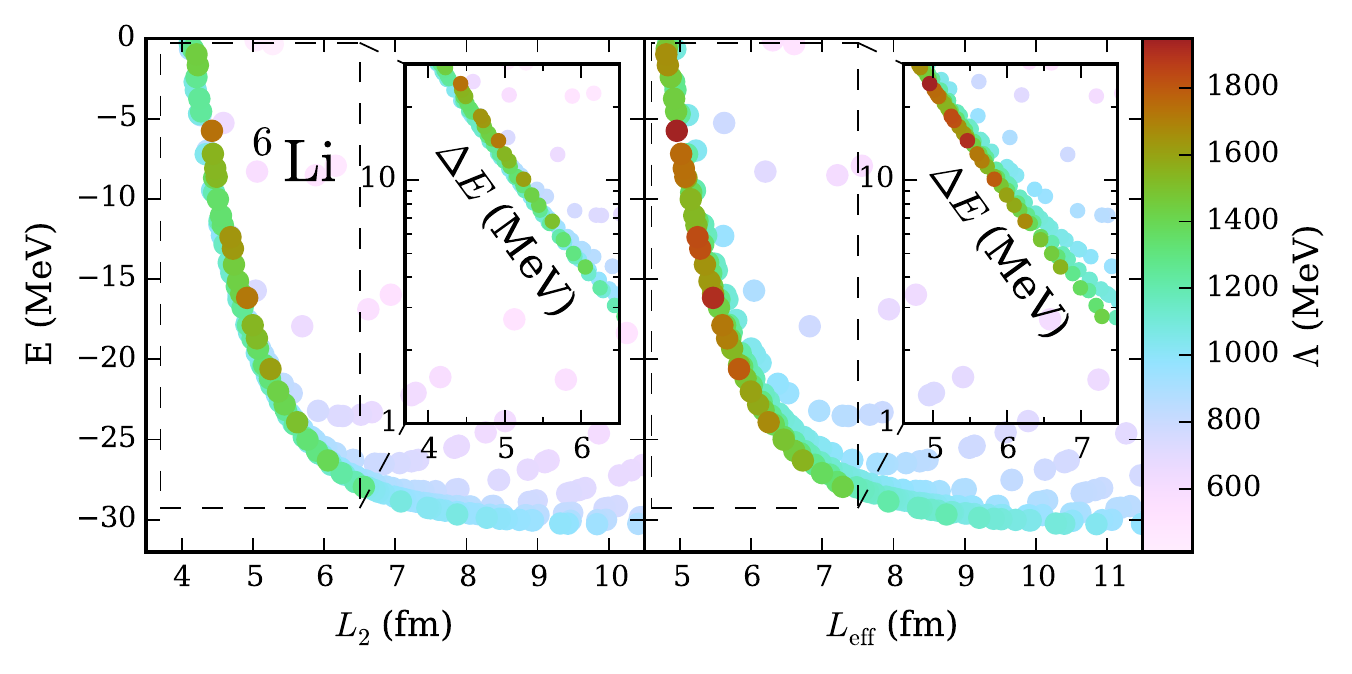}
  \caption{\co{} Ground-state energy of \nuc{6}{Li} plotted as a function of the
    IR scale determined by either $L_2$~\cite{more2013} (left panel) or \Leff{} from
    this work (right panel). The color of each circular marker
  indicates the UV cutoff of that calculation, with darker colors corresponding to larger cutoffs. 
  Insets show $\Delta E = E(L,\Lambda) -
    E_\infty$ on a semi-logarithmic scale.}  
  \label{fig:Li6Contrast}
\end{figure*}

The right panel of Fig.~\ref{fig:Li6Contrast} shows that a common, exponential
envelope is formed by the data with large UV cutoffs when plotted as a function
of \Leff{}. In particular, at a given value of \Leff{}, the energy with the
largest UV cutoff $\Lambda_\mathrm{eff}$ is lowest in energy. We also find that
results from smaller model spaces deviate rather quickly from the IR envelope
due to lack of UV convergence. The left panel of Fig.~\ref{fig:Li6Contrast}
shows the same energies plotted as a function of the IR scale $L_2$, with $N$ in
Eq.~\eqref{eq:L2} corresponding to the highest single-particle state in the
basis. While the points fall close to a line, no envelope is formed, and the
data with the highest UV cutoff $\Lambda_2=L_2/b^2$ is not lowest in energy at
given $L_2$. The comparison of the left and right panels demonstrates that
\Leff{} is a much more precise IR length for the \ncsm{} than $L_2$, and this
leads to more stable extrapolations. 

\rcsection{Extrapolation results}
The exponential IR extrapolations~\cite{coon2012,furnstahl2012,furnstahl2014}
can be generalized to the NCSM by employing the asymptotic wave function 
\be
\label{asympt}
\psi(\rho) \rightarrow e^{-\kappa\rho} -e^{-2\kappa{}L}e^{+\kappa{}\rho}
\ee
that is consistent with the Dirichlet boundary condition at hyper radius
$\rho=L$. Here, $\kappa$ denotes a hypermomentum. Asymptotically, i.e., for $\kappa{}L\to
\infty$, the approximation~(\ref{asympt}) holds for any value of the grand
angular momentum. For nonzero grand angular momentum, corrections to the
coefficient $\exp{(-2\kappa{}L)}$ are of order $(\kappa{}L)^{-1}$, similar to corrections due
to finite angular momentum for a single particle in three
dimensions~\cite{furnstahl2014}. Thus, we use a simple exponential form
\begin{equation}
  \label{extra}
  E(L) = E_\infty + a e^{-2\kappa{}_\infty L} \,,
\end{equation}
for IR extrapolations of bound-state energies. Here, the extrapolated energy
$E_\infty$ and the parameters $a$ and $\kappa{}_\infty$ will be fit to data points
obtained in model spaces characterized by the IR length $L$. For the two-body
system, $a$ and $\kappa{}_\infty$ are related to the asymptotic normalization
coefficient and binding momentum, respectively~\cite{more2013}.

In the NCSM the computational expense grows rapidly with increasing $N^{\rm tot}_{\rm max}$. The IR extrapolation of a bound-state energy is useful if the resulting $E_\infty$ (obtained from NCSM spaces with up to $N^{\rm tot}_{\rm max}$) is closer to the exact result than the variational
minimum energy that can be computed in a \ncsm{} space with $N^{\rm tot}_{\rm max}$. To locate the minimum, one needs at least three \ncsm{}
calculations. For IR extrapolations, one needs also at least three \ncsm{}
calculations, with parameters $\Leff$ and $\Lameff$ such
that (i) $\Leff$ significantly exceeds the radius of the nucleus under
consideration, (ii) $\Lameff$ significantly exceeds the UV cutoff of
the interaction, and (iii) the resulting energies are
negative.

\begin{figure*}[htbp]
  \centering
  \includegraphics[width=\textwidth]{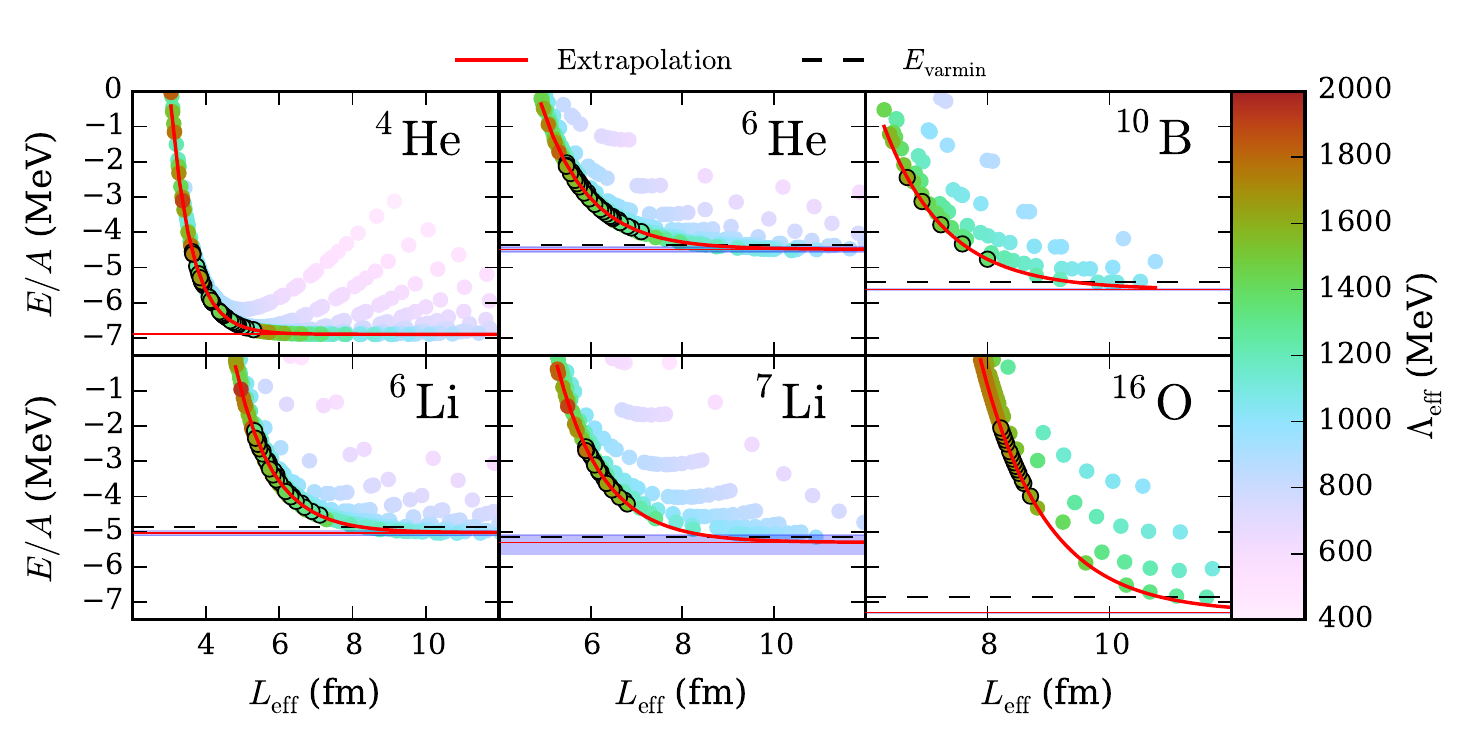}
  \caption{\co{} Extrapolations of the binding energy per particle for several
  $p$-shell nuclei computed with the \ncsm{}. The color of each circular marker
  indicates the UV cutoff of that calculation with darker colors corresponding to larger cutoffs.  Markers with a black border are included in the
  extrapolation.  The solid red (gray) curve shows the exponential
  fit~\eqref{extra}, and the horizontal red (gray) line marks the value of
  $E_\infty$ with uncertainty estimates indicated as blue (gray) bands.  The dashed black line marks the variational minimum $E_{\rm varmin}$ for the
  largest model space included in the fit.}
  \label{fig:Fits}
\end{figure*}

Figure~\ref{fig:Fits} shows extrapolations for the ground-state energies of
$\nuc{4}{He}$, $\nuc{6}{He}$, $\nuc{6}{Li}$, $\nuc{7}{Li}$,
$\nuc{10}{B}$, and $\nuc{16}{O}$.  For the $A=4, 6$ systems we can perform
\ncsm{} calculations in very large model spaces for which the ground-state
energies are virtually converged. However, in order to benchmark the
effectiveness of the extrapolation we artificially restrict our
data set to smaller models spaces, $\nmaxtot \le 10$ and
$\nmaxtot \le 14$ respectively.  For $\nuc{7}{Li}$, we use 
energies from model spaces with $\nmaxtot \le 17$. The extrapolations for the nuclei $\nuc{10}{B}$ and $\nuc{16}{O}$
employ energies from a single model space of $\nmaxtot=16$ and $\nmaxtot=20$,
respectively. Energies from smaller model spaces were not deemed sufficiently UV converged, as can be seen in the corresponding panels.
For each nucleus, we select data with $\Lameff$ large
enough that all points fall on a single narrow envelope and $\Leff$
large enough that $E(\Leff) - \Ei$ is exponential.  The blue (gray) horizontal bands give an estimate of the
uncertainly of the fit, obtained from refitting with all possible pairs of data excluded from the data set.
Table~\ref{tab:results} summarizes the results. The comparison to benchmark results shows that IR extrapolations are useful.

\begin{table}[hbt]
\caption{Energies (in MeV) of ground states with given spin and parity for several nuclei.  
  Benchmark results $E_{\rm ref}$ from coupled-cluster calculations for
  \nuc{16}{O}~\cite{ekstrom2013} and from the NCSM for 
  \nuc{4,6}{He} and \nuc{6}{Li} are obtained in large model spaces with
  $\nmaxtot = 20$ (equivalent to $\nmaxncsm=18$ for $A=6$). Extrapolated energies $E_\infty$ (with fit parameter $\kappa{}_\infty$ in units of fm$^{-1}$)
  and variational minimum energies $E_{\rm varmin}$ are from smaller model spaces with \nmaxtot{}.
\label{tab:results}} 
\begin{ruledtabular}
\begin{tabular}{c|dddcd} 
      Nucleus 
      & \multicolumn{1}{c}{$E_{\rm ref}$} 
      & \multicolumn{1}{c}{$E_{\rm var min}$}   
      & \multicolumn{1}{c}{$E_\infty$}
      & \multicolumn{1}{c}{\nmaxtot}   
      & \multicolumn{1}{c}{$\kappa{}_\infty$} \\ 
      \hline
\nuc{4}{He}$(0^+)$           & -27.76 &  -27.51 &  -27.59 & 10 & 0.87\\
\nuc{6}{He}$(0^+)$           & -27.13 &  -26.12 &  -26.92 & 14 & 0.49\\
\nuc{6}{Li}$(1^+)$           & -30.27 &  -29.18 &  -30.17 & 14 & 0.49\\
\nuc{7}{Li}$(\frac{3}{2}^-)$ &  $---$ &  -36.11 &  -37.14 & 17 & 0.50\\
\nuc{10}{B}$(3^+)$           &  $---$ &  -54.24 &  -56.29 & 16 & 0.50\\
\nuc{16}{O}$(0^+)$           & -130.1 & -109.77 & -116.75 & 20 & 0.47\\
\end{tabular}
\end{ruledtabular} 
\end{table}

\rcsection{Summary} We determined the IR length scale of the NCSM by
equating the intrinsic kinetic energy of $A$ fermions subject to the
NCSM truncation to the kinetic energy of $A$ fermions in a $3(A-1)$
hyper-radial well with Dirichlet boundary condition for the single
collective variable $\rho$. Calculations of \nuc{6}{Li} in large NCSM
spaces show that the resulting IR length \Leff{} is correctly
identified. We applied this result to extrapolate ground-state
energies in \nuc{4,6}{He}, \nuc{6,7}{Li}, \nuc{10}{B}, and
\nuc{16}{O}. The comparison with benchmark results shows that
extrapolated energies are closer to the benchmarks than the minimum
variational energies obtained in the model spaces utilized for the
extrapolation. Further progress would depend on a better understanding
of the extrapolation formula for the NCSM and in particular on
combined UV and IR correction terms. 

\begin{acknowledgments}
  We thank A.~Ekstr\"om, R.~J.~Furnstahl, S.~K\"onig, and P.~Maris for useful discussions,
   and H.~T.~Johansson and
  B.~D.~Carlsson for support with the large-scale \ncsm{}
  calculations. This material is based upon work supported by the
  U.S. Department of Energy, Office of Science, Office of Nuclear
  Physics under Awards No.\ DEFG02-96ER40963 (University of
  Tennessee) and No.\ DE-SC0008499 (NUCLEI SciDAC Collaboration) and under
  Contract No.\  DE-AC05-00OR22725 (Oak Ridge National Laboratory). It
  was also supported by the Swedish Foundation for International
  Cooperation in Research and Higher Education (STINT, IG2012-5158)
  and by the European Research Council under the European Community's Seventh
  Framework Programme (FP7/2007-2013) / ERC Grant Agreement
  No.\ 240603.
\end{acknowledgments}

\clearpage
\onecolumngrid
\section*{Supplementary Material}
\begin{table}[hbt]
\caption{Effective $\tilde{N}$ for \ncsm{}, needed for computing $L_{\rm eff} =
b\tilde{N}$ and $\Lambda_{\rm eff} = b^{-1}\tilde{N}$, where
$b=\sqrt{\hbar/(M\omega)}$ is the oscillator length.  For odd $N^{\rm tot}_{\rm
max}$ and even parity, use the value for $N^{\rm tot}_{\rm max}-1$, likewise for
even $N^{\rm tot}_{\rm max}$ and odd parity.  As an example, consider $^{6}$Li
computed with $\nmaxncsm=12$. In this case, $\nmaxtot = \nmaxncsm + A - 4=14$
and the parity is positive, therefore $\Leff=6.688\,b$.
\label{tab:neff}}
\newcommand{\mc}[1]{\multicolumn{1}{c}{#1}}
\newcommand{\ec}{\multicolumn{1}{c}{------}}

\begin{ruledtabular}
  \begin{tabular}{rrrrrrrrrrrrrrr} 
    & \multicolumn{13}{c}{$A$ $(\pi=+\null)$} \\ \cline{2-15}
    \mc{$N^{\rm tot}_{\rm max}$}
        & \mc{ 3} & \mc{ 4} & \mc{ 5} & \mc{ 6} & \mc{ 7} & \mc{ 8} & \mc{ 9} & \mc{10} & \mc{11} & \mc{12} & \mc{13} & \mc{14} & \mc{15} & \mc{16} \\
    \hline
     0  &   2.965 &   3.294 &     \ec &     \ec &     \ec &     \ec &     \ec &     \ec &     \ec &     \ec &     \ec &     \ec &     \ec &     \ec \\
     2  &   3.631 &   3.934 &   3.920 &   4.150 &     \ec &     \ec &     \ec &     \ec &     \ec &     \ec &     \ec &     \ec &     \ec &     \ec \\
     4  &   4.169 &   4.450 &   4.526 &   4.747 &   4.631 &   4.819 &     \ec &     \ec &     \ec &     \ec &     \ec &     \ec &     \ec &     \ec \\
     6  &   4.637 &   4.900 &   5.012 &   5.225 &   5.214 &   5.397 &   5.226 &   5.389 &     \ec &     \ec &     \ec &     \ec &     \ec &     \ec \\
     8  &   5.058 &   5.306 &   5.436 &   5.640 &   5.676 &   5.855 &   5.795 &   5.955 &   5.748 &   5.894 &     \ec &     \ec &     \ec &     \ec \\
    10  &   5.444 &   5.680 &   5.818 &   6.016 &   6.078 &   6.252 &   6.243 &   6.400 &   6.308 &   6.452 &   6.220 &   6.353 &     \ec &     \ec \\
    12  &   5.803 &   6.029 &   6.172 &   6.363 &   6.441 &   6.611 &   6.632 &   6.786 &   6.747 &   6.889 &   6.773 &   6.904 &   6.653 &   6.776 \\
    14  &   6.141 &   6.357 &   6.502 &   6.688 &   6.777 &   6.942 &   6.983 &   7.133 &   7.126 &   7.266 &   7.205 &   7.335 &   7.201 &   7.323 \\
    16  &   6.461 &   6.668 &   6.814 &   6.994 &   7.090 &   7.252 &   7.306 &   7.454 &   7.468 &   7.605 &   7.577 &   7.706 &   7.627 &   7.749 \\
    18  &   6.765 &   6.965 &   7.111 &   7.286 &   7.387 &   7.545 &   7.608 &   7.753 &   7.782 &   7.917 &   7.912 &   8.038 &   7.994 &   8.114 \\
    20  &   7.056 &   7.249 &   7.394 &   7.564 &   7.669 &   7.823 &   7.894 &   8.037 &   8.077 &   8.210 &   8.220 &   8.345 &   8.324 &   8.442 \\
    22  &   7.335 &   7.522 &   7.666 &   7.832 &   7.939 &   8.090 &   8.166 &   8.306 &   8.355 &   8.486 &   8.508 &   8.631 &   8.626 &   8.743 \\
    24  &   7.604 &   7.785 &   7.927 &   8.090 &   8.198 &   8.346 &   8.426 &   8.564 &   8.619 &   8.748 &   8.780 &   8.901 &   8.909 &   9.025 \\
    26  &   7.863 &   8.039 &   8.180 &   8.339 &   8.448 &   8.594 &   8.677 &   8.812 &   8.872 &   9.000 &   9.038 &   9.158 &   9.176 &   9.290 \\
    28  &   8.114 &   8.286 &   8.425 &   8.580 &   8.690 &   8.833 &   8.918 &   9.052 &   9.116 &   9.241 &   9.286 &   9.405 &   9.430 &   9.543 \\
    30  &   8.357 &   8.525 &   8.662 &   8.814 &   8.924 &   9.065 &   9.152 &   9.283 &   9.351 &   9.475 &   9.524 &   9.641 &   9.673 &   9.785 \\
    \hline\hline
    & \multicolumn{13}{c}{$A$ $(\pi=-\null)$} \\ \cline{2-15}
    \mc{$N^{\rm tot}_{\rm max}$}
        & \mc{ 3} & \mc{ 4} & \mc{ 5} & \mc{ 6} & \mc{ 7} & \mc{ 8} & \mc{ 9} & \mc{10} & \mc{11} & \mc{12} & \mc{13} & \mc{14} & \mc{15} & \mc{16} \\
     \hline
     1  &   3.190 &   3.489 &   3.755 &     \ec &     \ec &     \ec &     \ec &     \ec &     \ec &     \ec &     \ec &     \ec &     \ec &     \ec \\
     3  &   3.838 &   4.117 &   4.369 &   4.295 &   4.500 &     \ec &     \ec &     \ec &     \ec &     \ec &     \ec &     \ec &     \ec &     \ec \\
     5  &   4.360 &   4.622 &   4.862 &   4.887 &   5.086 &   4.940 &   5.114 &     \ec &     \ec &     \ec &     \ec &     \ec &     \ec &     \ec \\
     7  &   4.815 &   5.063 &   5.292 &   5.360 &   5.553 &   5.514 &   5.685 &   5.495 &   5.648 &     \ec &     \ec &     \ec &     \ec &     \ec \\
     9  &   5.226 &   5.461 &   5.680 &   5.771 &   5.958 &   5.969 &   6.136 &   6.058 &   6.209 &   5.989 &   6.129 &     \ec &     \ec &     \ec \\
    11  &   5.603 &   5.828 &   6.038 &   6.143 &   6.325 &   6.364 &   6.527 &   6.502 &   6.650 &   6.545 &   6.683 &   6.440 &   6.569 &     \ec \\
    13  &   5.955 &   6.171 &   6.373 &   6.486 &   6.663 &   6.720 &   6.879 &   6.886 &   7.031 &   6.981 &   7.116 &   6.991 &   7.117 &   6.857 \\
    15  &   6.287 &   6.494 &   6.689 &   6.807 &   6.979 &   7.049 &   7.205 &   7.231 &   7.374 &   7.357 &   7.490 &   7.420 &   7.545 &   7.403 \\
    17  &   6.600 &   6.800 &   6.989 &   7.110 &   7.278 &   7.356 &   7.509 &   7.550 &   7.691 &   7.694 &   7.826 &   7.790 &   7.913 &   7.828 \\
    19  &   6.900 &   7.092 &   7.276 &   7.399 &   7.563 &   7.647 &   7.797 &   7.848 &   7.986 &   8.006 &   8.135 &   8.121 &   8.244 &   8.192 \\
    21  &   7.186 &   7.373 &   7.551 &   7.675 &   7.835 &   7.924 &   8.071 &   8.129 &   8.266 &   8.296 &   8.424 &   8.427 &   8.547 &   8.519 \\
    23  &   7.461 &   7.642 &   7.815 &   7.940 &   8.096 &   8.189 &   8.333 &   8.397 &   8.532 &   8.571 &   8.697 &   8.712 &   8.831 &   8.820 \\
    25  &   7.726 &   7.902 &   8.071 &   8.195 &   8.348 &   8.443 &   8.585 &   8.654 &   8.786 &   8.833 &   8.957 &   8.981 &   9.099 &   9.101 \\
    27  &   7.981 &   8.153 &   8.318 &   8.442 &   8.592 &   8.689 &   8.828 &   8.901 &   9.031 &   9.083 &   9.206 &   9.237 &   9.354 &   9.365 \\
    29  &   8.229 &   8.397 &   8.558 &   8.681 &   8.828 &   8.926 &   9.063 &   9.139 &   9.267 &   9.323 &   9.445 &   9.482 &   9.598 &   9.617 \\
    31  &   8.470 &   8.633 &   8.790 &   8.913 &   9.058 &   9.156 &   9.291 &   9.369 &   9.496 &   9.556 &   9.675 &   9.718 &   9.832 &   9.858 \\
  \end{tabular}
\end{ruledtabular} 
\end{table}
 
\end{document}